\documentclass[doublecol]{epl2}

\title{Lattice anomalies in the FeAs$_{\rm 4}$ tetrahedra of the
NdFeAsO$_{\rm 0.85}$ superconductor that disappear at T$_{\rm c}$}

\author{M. Calamiotou \inst{1}, I. Margiolaki \inst{2}, A. Gantis \inst{1}, E. Siranidi \inst{3},
Z.A. Ren \inst{4}, Z.X. Zhao \inst{4}, and E. Liarokapis \inst{3}}
\shortauthor{M. Calamiotou \etal}

\institute{
  \inst{1} Solid State Physics Department, School of
Physics, University of Athens, GR-15784 Athens, Greece\\
  \inst{2} ESRF, BP 220, Grenoble, Cedex 9, F-38043 France\\
  \inst{3} Department of Physics, National Technical University of Athens,
157 80 Athens, Greece\\
  \inst{4} National Laboratory for Superconductivity, Institute of
Physics and Beijing National Laboratory for Condensed Matter
Physics, Chinese Academy of Sciences, P.O.Box 603, Beijing, China}


\abstract{High resolution synchrotron X-ray powder diffraction
(SXRPD) was used to study the temperature dependence of the oxygen
deficient NdFeAsO$_{0.85}$ superconducting compound. By employing
a dense temperature sampling we have managed to reveal unnoticed
structural modifications that start around $\sim$180K, and
disappear at the transition temperature. The data show minor
changes of the structural characteristics in the Nd-O charge
reservoir layer while in the superconducting Fe-As layer the
FeAs$_{4}$ tetrahedron shows gradual modifications below
$\sim$180K, which suddenly disappear at T$_{\rm c}$ strongly
indicating a connection with superconductivity.}

\begin{document}

\maketitle

\section{Introduction}

The discovery of superconductivity in iron- based layered
compounds REFeAsO$_{\rm 1-x}$F$_{\rm x}$ (RE=Sm, Nd, Ce, Pr, Gd)
belonging to the family of oxypnictides \cite{Kamihara} attracted
a lot of experimental and theoretical attention. However the
mechanism, which induces superconductivity by doping the parent
REFeAsO compound, is still controversial. REFeAsO has been found
to undergo a structural tetragonal-to-orthorhombic phase
transition upon cooling \cite{Nomura,Margadonna,Fratini} and
exhibits a spin density wave antiferromagnetic (AF) ordering
\cite{Nomura,Dong,Cruz,Chen} closely resembling the cuprates. Both
phenomena seems to be suppressed by doping and with the appearance
of superconductivity. However in the hole-doped Nd$_{\rm
1-x}$Sr$_{\rm x}$FeAsO superconducting compound the structural
phase transition is not suppressed by increasing the Sr$^{\rm 2+}$
content \cite{Kasperkiewicz}. In the oxygen deficient
superconducting compound  REFeAsO$_{{1-\delta}}$ \cite{Ren} with
T$_{\rm c}$ ranging from 31.2K for RE=La to 55K for RE=Sm and
$\delta$=0.15, the tunable oxygen content leads to the occurrence
of superconductivity strongly resembling the situation in cuprate
superconductors.

Low temperature Fourier Transform Infrared (FTIR) studies in one
of the highest T$_{\rm c}$ (53.5K) pnictides (NdFeAsO$_{\rm 0.85}$
\cite {Ren}) have revealed unexpected spectral modifications and
temperature dependent anomalies for certain modes at $\sim$180K
\cite{Siranidi}.  Some marginal evidence of inflection points near
T$_{\rm c}$ in the temperature dependence of the bond angle
$\alpha_{\rm As-Fe-As}$ and bond length d$_{\rm Fe-As}$ has been
recently reported for a NdFeAsO$_{\rm 0.85}$ sample \cite{Kumai}.
In addition a femtosecond spectroscopy study of a single crystal
SmFeAsO$_{0.8}$F$_{0.2}$ superconducting sample revealed the
presence of a \emph{pseudogaplike} feature with an onset above
180K \cite{Mertelj}. Moreover, in the non-superconducting
LaFeAsO$_{\rm 1-x}$F$_{\rm x}$ a clear lattice anomaly has been
detected at $\sim$180K that disappears upon AF ordering
\cite{Qureshi}. Detailed and accurate structural studies as a
function of temperature are therefore essential in order to
establish a clear picture of any correlation between crystal and
electronic structures in the iron based superconductors. We
present here high resolution synchrotron powder diffraction
results as a function of temperature from the superconducting
NdFeAsO$_{\rm 0.85}$ system (T$_{\rm c}$=53.5K) \cite{Ren}. Our
data recorded with a dense sampling in the temperature range down
to 10K provide clear evidence of subtle lattice distortions that
start at T$_{onset}$$\approx$180K and disappear at T$_{c}$.

\section{Experimental}

We have collected SXRPD data on beamline ID31 at the European
Synchrotron Radiation Facility (ESRF), Grenoble, France. The
experimental set-up is described in detail elsewhere \cite{Fitch}.
The short wavelength ($\lambda =0.39998(5)$\AA), to reduce
absorption, was selected with a double-crystal Si(111)
monochromator and calibrated with Si NIST standard. Optimum
transmission was achieved by enclosing the finely ground sample in
a 0.6mm diameter borosilicate glass capillary and appropriate
spinning of the capillary in the beam ensured for a good powder
averaging. An exchange gas continuous liquid-helium flow cryostat
with rotating sample rod was used to cool the sample down to 10K.
High statistics high resolution diffraction patterns
(2$\theta=1-53^o$, d spacing 22.9\AA-0.45\AA, 0.004$^{\rm o}$
step, scan time 30min) have been recorded at selected
characteristic temperatures. High resolution diffraction patterns
(short scans with 2$\theta =1-33^{\rm o}$, d spacing
22.9\AA-0.7\AA, 0.004$^{\rm o}$ step and scan time 1min), were
also collected upon heating (thermodiffractograms) from 10K up to
295K in order to follow the evolution of the lattice constants and
interatomic distances with a dense temperature sampling. Due to
the short time interval of the short scans upon heating the c-axis
values show that there was an $\sim$10K difference between the
nominal and the actual value of temperature. Therefore, all
temperature values of the short scans are shifted in the following
plots by this amount to lower values.

Data analysis has been performed with the GSAS+EXPGUI suite of
Rietveld analysis programs \cite{Larson} assuming the tetragonal
(SG P4/nmm) structure for the NdFeAsO$_{0.85}$ phase. The
pseudo-Voigt function corrected for asymmetry owing to axial
divergence and anisotropic broadening was used for the peak
profile(profile No 4 in GSAS\cite{Stephens}). The data have been
sequentially refined starting from the diffraction pattern
recorded at 10K. The high statistics diffraction patterns,
collected at the characteristic temperatures, have been used to
fully refine the values of the profile parameters, the Nd and As
coordinates and the atomic displacement factors U$_{\rm iso}$ of
NdFeAsO$_{0.85}$. Oxygen occupancy has been fixed to the nominal
value, justified by the agreement of the Nd-As bond length value
to that obtained from a neutron diffraction study in a series of
variable oxygen stoichiometry NdFeAsO$_{1-y}$ samples \cite{Lee}.
The refinements take into account small amounts of the impurity
phases FeAs (5$\pm$0.9\%wt) and Nd$_{2}O_{3}$ (1.4$\pm$0.3\%wt)
while a few very weak peaks due to other minor impurity phases
($<0.1$\%wt) have been excluded. We note that excluding also the
peaks of the two main impurity phases does not affect the
structural results.

\section{Results and discussion}

Figure 1 shows the diffraction pattern at 10K together with the
result of the Rietveld refinement. The high resolution data reveal
no splitting or broadening of the 110 reflection profiles that
would be the sign of a tetragonal-orthorhombic (T-O) phase
transition down to 10K, as shown in the inset in Fig.1. Table I
shows the structural parameters at different temperatures obtained
from the Rietveld refinement of the high statistics patterns. The
position of the As atom does not change in the whole temperature
range examined. The fractional coordinate of the Nd atom along the
c-axis remains constant down to 140K, it jumps to a higher value
within 140-60K, indicating that the Nd atom shifts closer to the
superconducting Fe-As layer, and then it remains practically
constant to the lowest temperature studied. The
thermodiffractograms between the characteristic temperatures have
been analyzed using both the Rietveld and the Lebail (whole-powder
pattern decomposition method not based on the structural model)
method. Minor impurity peaks were excluded from the refinement.
The Rietveld analysis has been performed combined with the high
statistics data, i.e., the atomic positions and the atomic
displacement factors U$_{\rm iso}$ were fixed to the corresponding
values obtained from the refinements of the high statistics data.
In the temperature region 60K-140K, where the fractional
coordinate of the Nd atom has been found to exhibit a change, we
have assumed a smooth linear change of $\emph{z}$. The LeBail
refinement, not based on a specific structural model, resulted to
identical values of the lattice constants.

The temperature dependence of the c- and a-axis (Fig.2) reveals
two important features: a slope change at T$\sim$180K and a subtle
anomaly at T$_{\rm c}$. The lattice of NdFeAsO$_{0.85}$ contracts
anisotropically, the relative change being $\Delta$a/a=0.136\% and
$\Delta$c/c=0.37\% from 295K to 60K, in agreement with previous
reports \cite{Kumai}. The distance between two adjacent Nd and As
atoms that reflects the spacing between the charge reservoir Nd-O
and the superconducting Fe-As layer (Fig.3a), decreases from 295K
to 60K by 0.235\% following the reduction of the c-axis. The Nd-O
bond contracts upon cooling (0.159\%) following the contraction of
the a-axis (Fig.3b), while the Nd-O-Nd bond angle exhibits only
marginal changes with temperature (Fig.3c). On the contrary, the
Fe-As bond length in the superconducting layer exhibits a bigger
change (0.235\%) comparable to that of the c-axis (Fig.4a). The
tetrahedral bond angle $\alpha_{\rm As-Fe-As}$ (Fig.4b) is
increasing while $\beta_{\rm As-Fe-As}$ (Fig.4c) is decreasing,
indicating an increase of tetrahedral distortion with decreasing
temperature. The FeAs$_{4}$ coordination tetrahedron is distorted
from the ideal symmetry. Robinson $\emph{et al}$\cite{Robinson}
have found that a quantitative measure of the distortion in such a
case is the variance of the tetrahedral angle expressed as
$\sigma^{2}= \displaystyle\sum_{\rm
i=1}^{6}(\theta_{i}-109.47^{o})^{2}/5$ where $\theta_{i}$ are the
bond angles of the distorted tetrahedron and $109.47^{o}$ the
corresponding one  in an ideal tetrahedron. Fig.5a shows the
distortion of the Fe coordination tetrahedron and Fig.5b presents
the temperature evolution of the volume \cite{Finger} of the
FeAs$_{4}$ tetrahedra in comparison to that of the cell volume
both normalized to their values at 295K.

The temperature evolution of all structural features (Figs.2-5)
reveals anomalies at characteristic temperatures. Specifically, at
T$_{onset}$$\sim180$K a change of slope is evident in the lattice
constants (Fig.2), the inter-layer Nd-As distance (Fig.3a), the
intra-layer Nd-O bond length (Fig.3b) and angle (Fig.3c), as well
as in the geometrical characteristics of the FeAs$_{4}$ tetrahedra
(Figs.4,5). This change is much more pronounced in the
superconducting Fe-As layer. Furthermore, around
T$_{f}$$\sim$135K, a modification in the temperature dependence is
evidenced for all bond length characteristics (Figs.3-5). It looks
like a new order parameter coupled to the lattice and mostly to
the Fe-As tetrahedra sets in at $\rm T_{onset}$ which is completed
at $\rm T_f$ (Figs.4-5). At T$_{c}$ the Fe-As bond length and the
distortion of the FeAs$_{\rm 4}$ tetrahedra exhibit a sudden
change (Figs.4-5) while the Nd-O bond length and angle values
saturate (Fig.3). The modifications around T$_{\rm c}$ are not so
pronounced in the charge reservoir planes being less sensitive to
changes with temperature. The structural data indicate a bigger
effect in the superconducting Fe-As planes. The relative
contraction of the FeAs$_{\rm 4}$ volume deviates in a profound
way from that of the unit cell below $\rm T_{onset}$ where the
angle distortion is also modified. Crossing T$_{\rm c}$ the
anomalies are obviously reduced (Fig.5).

The lattice distortions  appear below $\sim$180K where certain IR
modes exhibit an anomalous behavior in the same compound
\cite{Siranidi}. The non-superconducting undoped NdFeAsO compound
is known to exhibit a tetragonal to orthorhombic phase transition
that starts at 150K \cite{Fratini}, and a magnetic ordering of the
iron spins below $\sim$141K \cite{Chen}. One could assume that the
lattice anomaly observed in the region 180K-T$_{c}$, originates
from a T-O structural phase transition because such a phase
transition can be traced in other superconducting oxypnictides
\cite{Margadonna,Kasperkiewicz}. However our high resolution
diffraction data (inset in Fig.1) do not support such possibility.
On the other hand the iron spins could order antiferromagnetically
along the c-axis at $\sim$141K \cite{Chen}, but this phase should
be absent in the optimally doped compound, though it may be
present at low doping levels \cite{Rotter,Park}. The assumption of
two chemically separated phases cannot be justified from the XRD
results since no splitting or broadening of the xrd peaks was
observed. However, a mesoscopic phase separation cannot be
excluded, at least in the underdoped pnictides \cite{Park}. The
observation of similar lattice distortions that start below
$\sim$180K and relaxes at T$_{\rm N}$ in the non-supeconducting
LaFeAsO$_{\rm 1-x}$F$_{\rm x}$ \cite{Qureshi} and the appearance
of a new photo-induced reflectivity component in the
superconducting SmFeAsO$_{\rm 0.8}$F$_{\rm 0.2}$ \cite{Mertelj}
point to a lattice distortion effect of common origin in all these
pnictides. But while in the non-superconducting compound with the
AF ordering the effect disappears at T$_{\rm N}$, in the
photo-induced reflectivity results that probe the electronic
states the component remains into the superconducting state, and
in our structural data of superconducting compound the lattice
distortions are relaxed at T$_{\rm c}$. It appears that another
order parameter, which is common to all compounds sets-in around
T$_{\rm onset}$ and affects spin, lattice, and electronic states.

Based on the Fe-As$_{\rm 4}$ angle modifications (Fig.5a) one
could assume local distortions that involve orbital ordering in
the Fe planes. Martinelli et al \cite{Martinelli} have proposed
that the structural transition could originate by this distortion
that brings a similarity of the oxypnictide systems to the
manganites. At $\rm T_{onset}$ both the volume and the distortion
of the FeAs$_{4}$ tetrahedra show an anomaly that saturates at
$\rm T_f$ and disappears below T$_{c}$. The direct effect on the
lattice and more precisely the anomalous contraction of the volume
of the FeAs$_{\rm 4}$ tetrahedra (Fig.5b) and the increase of
their distortion indicates a polaronic mechanism. In such a case,
the polarons should start been formed around 180K. In the undoped
compound the polarons could be related with the AF ordering
\cite{Qureshi}. In the presence of carriers, the AF ordering is
suppressed and the instability remains down to T$_{\rm c}$
(Fig.5b). The return of the volume of the FeAs$_{\rm 4}$
tetrahedron (Fe-As bond length) below T$_{\rm c}$ to the
anticipated value from the high temperature data could be due to
the delocalization of the polarons in the superconducting phase.
Whatever happens our data show that there are modifications of the
c-axis at $\rm T_{onset}$ and T$_{c}$ (Fig.2), and indicate that
the lattice effects occur mainly within the FeAs$_{\rm 4}$
tetrahedra and are related with the carriers and possibly with
superconductivity. Anomalies observed at similar temperatures
indicate that this might be a general feature of these compounds
and reserve further investigation on all pnictides.

In conclusion, the high resolution synchrotron diffraction data
collected with dense sampling in the temperature range 10-295K
revealed the presence of subtle lattice anomalies, which coincide
with those in the IR modes at $\sim$180K and disappear at the
critical temperature. These lattice anomalies, apparently not
connected with a structural phase transition, are more evident in
the supercconducting Fe-As planes consisting of a distortion and
contraction of the FeAs$_{\rm 4}$ tetrahedra and a reduction of
the Nd-As distance indicating charge transfer. The disappearance
of the anomaly across T$_{\rm c}$ points to a connection to
superconductivity.

\acknowledgments {We thank the ESRF for providing synchrotron beam
time at ID31 instrument and Dr. Larry Finger for his advice with
respect to the use of program DRAWxtl.}

\bigskip

\section {FIGURE CAPTIONS}

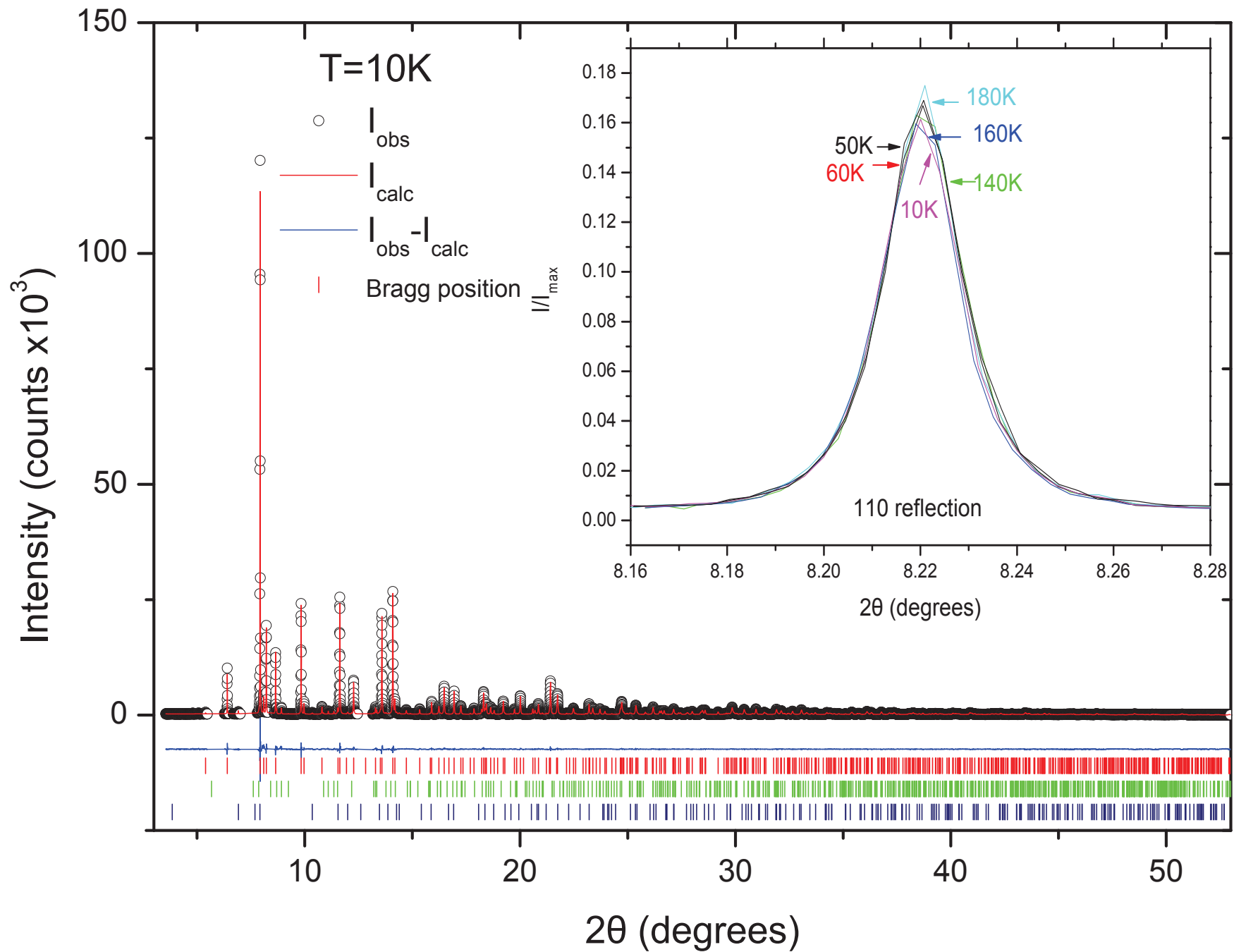
Figure 1. Experimental (circles), calculated (continuous line)
SXRPD patterns and their difference (bottom line) at T=10K. Bars
indicate the theoretical Bragg peak positions for
NdFeAsO$_{0.85}$. The two lower rows of bars indicate the Bragg
peak positions of the impurity phases FeAs and Nd$\rm _{2}O_{3}$
respectively. The inset shows the profiles of the 110 reflection
at different temperatures. Intensities are scaled to the most
intense line in the spectrum Imax and peaks are shifted to be
superimposed.

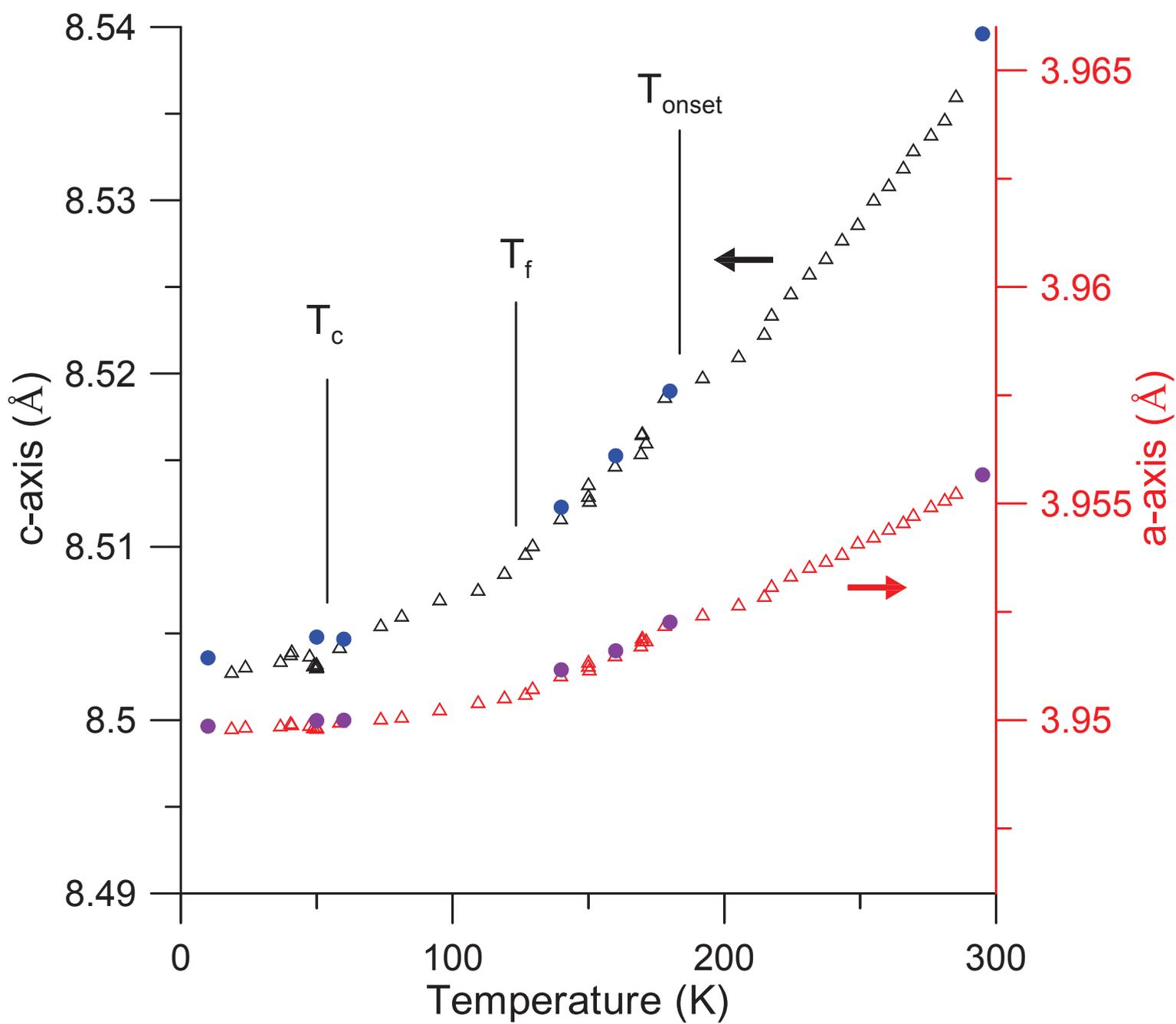
Figure 2. The dependence of the c- and a-axis on temperature. Full
circles: high statistics scans. Open triangles:
thermodiffractograms. Errors are smaller than symbols.

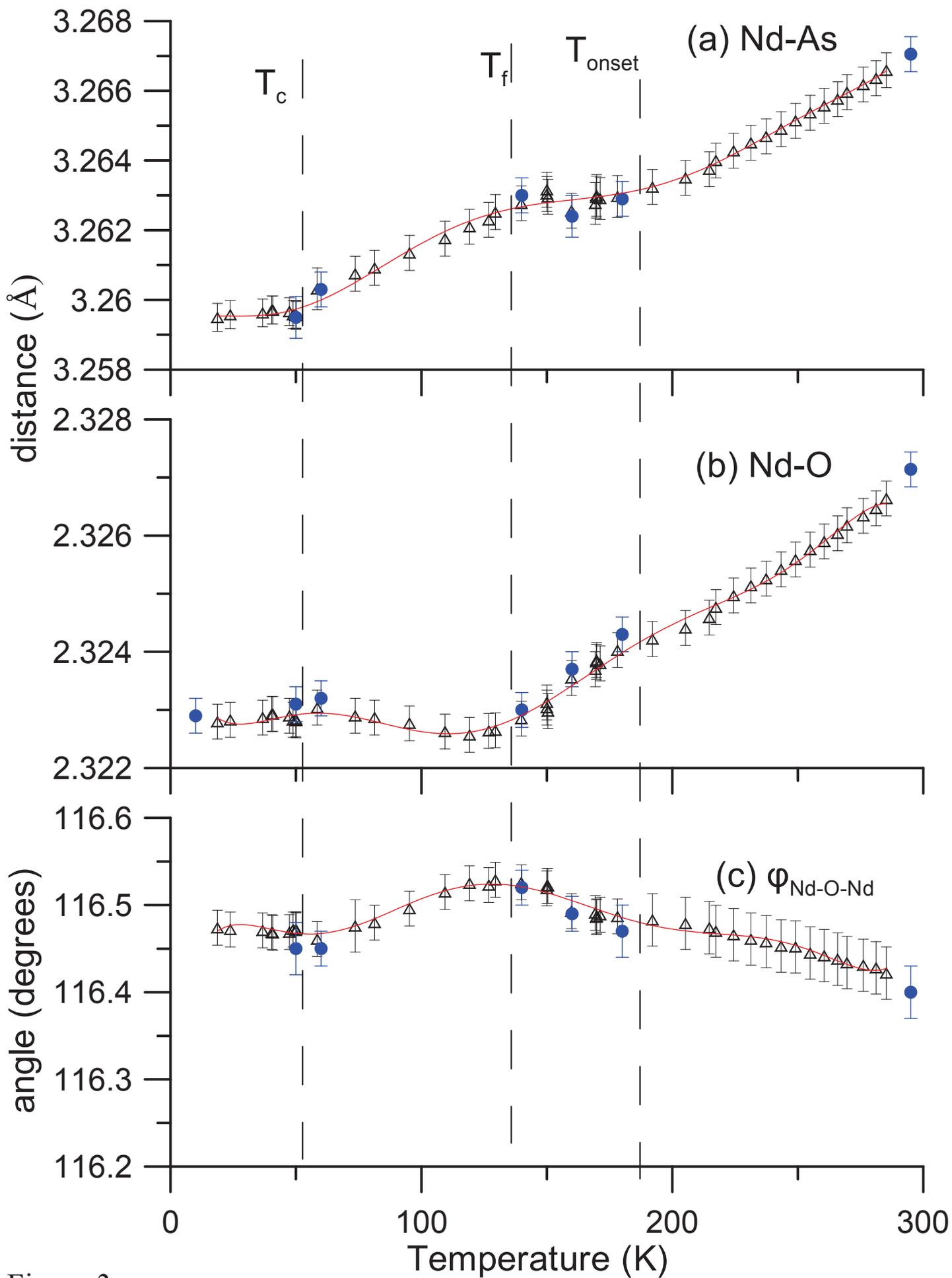
Figure 3. The change of (a) the inter-layer distance between two
adjacent Nd and As atoms, (b) the intra-layer Nd-O bond length,
and (c) the Nd-O-Nd angle with temperature. Full circles: high
statistics scans with estimated standard deviations (esd's). Open
triangles: thermodiffractograms with error bars calculated on the
basis of error propagation from the corresponding errors of
lattice constants and atomic positions. Lines are polynomial fits
to guide the eye.

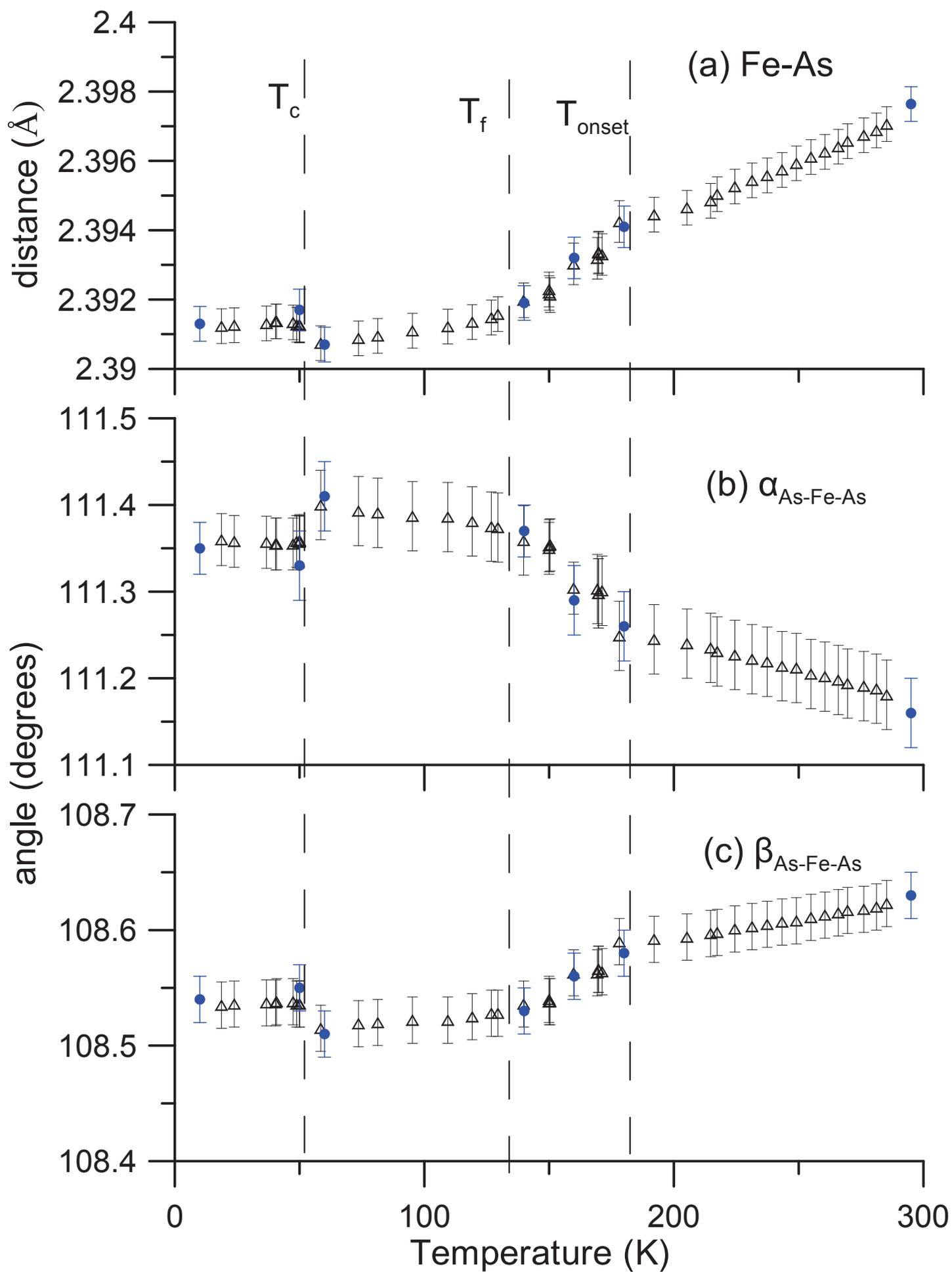
Figure 4. Temperature evolution of the intra-layer Fe-As bond
length (a) and Fe-As-Fe angles (b) and (c). Full circles: high
statistics scans with esd's. Open triangles: thermodiffractograms
with error bars calculated on the basis of error propagation from
the corresponding errors of lattice constants and atomic
positions.

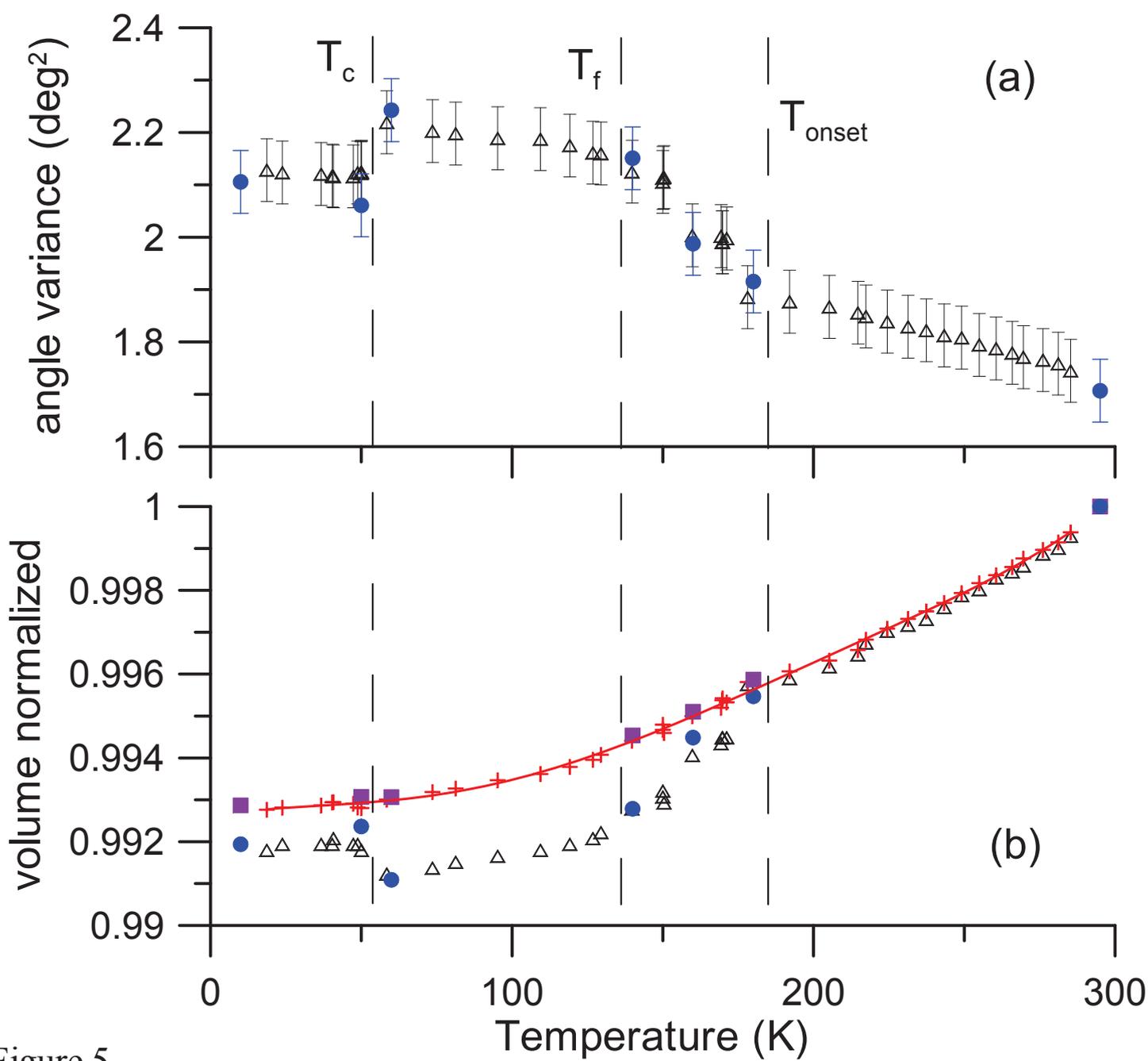
Figure 5. Temperature evolution of (a) the tetrahedral distortion
(angle variance), (b) the normalized volumes of FeAs$_{\rm 4}$
(circles and triangles) and unit cell (squares and crosses) to
their values at 295K. Full circles: high statistics scans with
esd's. Open triangles: thermodiffractograms with error bars
calculated on the basis of error propagation from the
corresponding errors of lattice constants and atomic positions.
\bigskip

\bigskip

\begin{table*}
\caption{\label{tab:table1}Rietveld refinements of synchrotron
diffraction data for NdFeAsO$_{0.85}$ at different temperatures.
The space group is P4/nmm with Nd on
2c($\frac{1}{4}$,$\frac{1}{4}$,z), Fe on
2b($\frac{3}{4}$,$\frac{1}{4}$,$\frac{1}{2}$), As on
2c($\frac{1}{4}$,$\frac{1}{4}$,z), O on
2a($\frac{3}{4}$,$\frac{1}{4}$,0).}
\begin{center}
\begin{tabular}{llllllllll}
                                         & & T=10K       & T=50K       &  T=60K      &  T=140K     &  T=160K     &  T=180K                \\
\hline
 Nd    &$\emph{z}$                        & 0.14381(6)  & 0.14383(7)  & 0.14385(6)  & 0.14356(6)  & 0.14363(7)  & 0.14364(7)             \\
       &U$_{iso}$(x100$\mathrm{\AA}^{2})$& 0.144(6)    &  0.131(7)   & 0.145(6)    & 0.279(7)     &  0.314(8)   & 0.293  (8)              \\
 Fe    &U$_{iso}$(x100$\mathrm{\AA}^{2})$& 0.18(2)     & 0.16(2)     & 0.26(2)     & 0.39(3)      & 0.39(3)     & 0.40(3)                  \\
 As    &$\emph{z}$                       &  0.6586(1)   & 0.6586(1)   & 0.6584(1)   & 0.6584(1)   & 0.6586(1)   & 0.6587(1)                 \\
       &U$_{iso}$(x100$\mathrm{\AA}^{2})$& 0.18(1)     & 0.16(2)     & 0.26(1)     & 0.38(1)     & 0.40(2)     & 0.40(1)                  \\
 O     &U$_{iso}$(x100$\mathrm{\AA}^{2})$& 1.2(2)      & 1.6(3)      & 1.7(2)      & 1.5(2)      & 1.4(2)      & 1.4(2)                     \\
 &a ($\mathrm{\AA}$)                      & 3.94986(1)  & 3.94999(1)  & 3.95000(1)  & 3.95116(1)  & 3.95160(1)  & 3.95226(1)                  \\
 &c ($\mathrm{\AA}$)                      & 8.50360(4)  & 8.50479(4)  & 8.50467(4)  & 8.51228(5)  & 8.51525(5) & 8.51899(5)                  \\
 &R$_{wp}$                                & 0.1094      & 0.1410      & 0.1357      & 0.1183      & 0.1209      & 0.1322            \\
 &R(F$^{2}$)                              & 0.0951      & 0.1050      & 0.0817      & 0.884      & 0.0807       & 0.949          \\

\end{tabular}
\end{center}
\end{table*}

\end{document}